\documentclass[journal]{IEEEtran}
\usepackage[noadjust]{cite} % This is for having a range of references; for example [3]-[12]
\usepackage{amsmath}
\usepackage{array}
\usepackage{graphics}
\usepackage[pdftex]{graphicx}
\usepackage{amssymb}
\usepackage{amsfonts}
\usepackage{nicefrac}
\usepackage{color}
\usepackage[dvipsnames]{xcolor}
\usepackage{tikz}
\usepackage{soul}

\def\1{{\mathbf{1}}}

\DeclareMathOperator*{\diag}{diag}

\begin{document}
\title{On the search for expanded grid control capabilities: Discrete control on emerging power technologies}
\author{H\'{e}ctor Pulgar-Painemal, \IEEEmembership{Senior Member, IEEE}, Sebastian Martinez-Lizana, \IEEEmembership{Graduate Student Member, IEEE}
\thanks{This material is based upon work supported by the National
Science Foundation under NSF CAREER Grant No. 2044629. This work also
made use of shared facilities
supported by the Engineering Research Center Program of
the National Science Foundation and the Department of Energy
under NSF Award No. EEC-1041877 and the CURENT
Industry Partnership Program.}
\thanks{Pulgar-Painemal and Martinez-Lizana are with the Department of Electrical Engineering and Computer Science, University of Tennessee, Knoxville, TN 37996, USA. e-mail: hpulgar@utk.edu, smart118@vols.utk.edu}}

\maketitle

\begin{abstract}
This letter proposes discrete changes in the power output of emerging power technologies (EPT) for controlling oscillations and frequency excursions. For the former, a new perspective is proposed that connects oscillations with the transient shift of the system equilibrium point. This is transformative as discrete control can be applied to multi-modal systems for the first time, without any model aggregation. For the latter, new insights are provided in regard to the nature of the discrete actions. Applications to a 2-bus, 9-bus and 39-bus test systems are presented. Through the proposed scheme, EPT can be enabled with controls that recognize their characteristics, while expanding grid dynamic capabilities with the addition of new effective controllers.
\end{abstract}

\section{Introduction}
Due to the decommissioning of old conventional power plants and projected higher penetration levels of non-conventional renewable sources (NCRS), frequency excursions will be more severe and the risk for more recurrent appearances of poorly damped oscillations will be higher. The former is related to a relative reduction of system inertia. The latter has a more complex nature and depends on multiple factors. Inertia distribution and grid topology have been found to be two of the most relevant factors involved with the appearance of critical oscillations \cite{Pulgar1,Wang_Yajun}. In addition, these can be worsened by the NCRS deployment away from load centers, redirection of power flows, and the dynamic interactions among synchronous generators (SGs) and NCRS. If these factors are negatively affected along the grid evolution towards higher NCRS penetration levels, oscillations will be a seriously critical and recurrent problem in the forthcoming grid. Several control schemes have been proposed to tackle these issues, such as synthetic inertia, damping controllers, or virtual synchronous machines---all of which are continuous in nature. Although fast and effective, these controllers, however, are significantly limited by operational constraints when implemented in wind turbines (WTs) and PV solar plants---this is due to the adoption of maximum power point tracking (MPPT) logic in control schemes. To overcome this obstacle, power curtailment has been used for allowing room to continuously modulate up and down their power output. Any profit reduction due to this planned energy spillage can be seen as an indirect control cost, and it can be excessively high, especially with increasing NCRS penetration levels. When implemented in other components such as energy storage systems (ESS), these controllers are less restrained and can perform to their full potential. Still, ESS installed power is not significant yet in most of the grids. Moreover, the most common type are batteries, which suffer from very limited maximum full depth charge/discharge cycles; if batteries are used significantly, their lifespan can be negatively affected.

Oscillations can be tackled through power system stabilizers (PSS) installed in selected SGs, but they are not always effective \cite{McCalley}, and this may be exacerbated in the future grid with features that can facilitate the appearance of critical oscillations. Regarding frequency excursions, a recent report \cite{Eto} urges the enabling of all generators for primary frequency response to the extent feasible and recognizes that further controls such as synthetic inertia in converter-interfaced components are not sustained for effective regulation. This letter proposes discrete changes in the power output of EPT for increased grid control capabilities regarding oscillations and frequency excursions. The initial results of a long-term project on this subject are presented: (a) connection of the system equilibrium point with the oscillation orbits that allows applying discrete control in multi-modal systems for the first time, and (b) new insights about the nature of the discrete actions for frequency excursions. The authors hope that this letter encourages further research on this area, as the grid can be greatly benefited from new effective controllers in EPT that recognize their particular characteristics and operational constraints.

\section{Discrete control}
In the case of WTs and PV solar plants, a discrete logic enable them for effective control by a transient stepwise reduction of their power output---without the need for curtailment. In the case of batteries, intuitively, lesser power usage for control purposes due to an effective discrete logic may help minimize any negative impact on the lifespan of batteries. Discrete control and the Pontryagin's maximum principle have been extensively applied in other fields such as aerospace \cite{Trelat}. However, when it comes to multi-modal systems such as the electric power grid, plus the inclusion of a high number of controllable components (CCs), the solution of a discrete control formulation as known can become quite cumbersome. A new perspective is needed.

\subsection{Discrete electromechanical oscillation control (DEOC)}
Moved by the effectiveness of discrete control, much work was done on this DEOC problem a few decades ago \cite{Kosterev2,Chang1}. To deal with the multi-modal nature of electric power systems, several simplifying assumptions were made to make the problem more tractable such as focusing on a single inter-area oscillation mode, system aggregation on both of the oscillation ends, and a unique CC located right on the oscillation path---typically the reactance in series compensated lines. Moreover, because a closed-form optimal switching function could not be found, numerical solutions were obtained by solving the time inverse problem \cite{Chang1}, which suffered from instabilities. Although conceptually interesting, these research efforts lacked practicality as it would be unlikely that a grid disturbance would excite only a single oscillation mode. This letter presents new findings that enable DEOC for multi-modal systems and provides a closed-form optimal switching function for the first time. This is possible by offering a new perspective that connects the system's oscillatory behavior with the transient shifting of the equilibrium point. For illustrative purposes, consider a system with $n$ buses, $n_g$ SGs represented by a classical model, and $n_c$ CCs connected in some of the $n-n_g$ non-generator buses:

{\small
\begin{align}
\underbrace{\left[\begin{array}{c}
\dot{\delta}\\
\dot{\omega}\\
\end{array}\right]}_{\dot{x}}&=\underbrace{\left[\begin{array}{cc}
0 & \omega_s I_{n_g}\\
-\frac{1}{2}H^{-1}B_a & 0\\
\end{array}\right]}_{A} \underbrace{\left[\begin{array}{c}
\delta\\
\omega\\
\end{array}\right]}_{x}+\left[\begin{array}{c}
-\omega_s \1_{n_g}\\
h\\
\end{array}\right]\label{eq:Part3_StateSpaceModel}
\end{align}}
where $h=\frac{1}{2}H^{-1} \big[ P_m+B_bP_L-B_c P_0-B_c \Delta P (\mu_{t_{on}}(t)-\mu_{t_{off}}(t)) \big]$; $\delta$ rad, $\omega$ p.u., and $P_m$ p.u. $\in \mathbb{R}^{n_g}$ are the vectors of loading angle, speed and mechanical power of the SGs; $P_L \in \mathbb{R}^{n-n_g}$ p.u. is the power load vector; $\omega_s=120 \pi$ rad/s; $P_0 \in \mathbb{R}^{n_c}$ p.u. the CCs initial injected power vector; $\Delta P \in \mathbb{R}^{n_c}$ p.u. the CCs power change vector; $\mu_{\tau}(t)$ the Heaviside step function at time $\tau$; $t_{on}, t_{off}$ s switching times---with $t_{on}<t_{off}$; $H=\diag \{H_1,...,H_{n_g}\} \in \mathbb{R}^{n_g \times n_g}$ s the SGs inertia matrix; $I_{n_g}$ the ${n_g \times n_g}$ identity matrix; and $\1_{n_g}$ an $n_g \times 1$ vector with all its components being 1. The matrices $B_a \in \mathbb{R}^{n_g \times n_g}$, $B_b \in \mathbb{R}^{n_g \times (n-n_g)}$ and $B_c \in \mathbb{R}^{n_g \times n_c}$ are obtained using the dc load flow formulation and by eliminating algebraic variables. Note that the system behaves as an undamped harmonic oscillator. If $t<t_{on}$ or $t>t_{off}$, when DEOC is off, the system equilibrium point is defined as  $x_e=[\delta_e~~\omega_e]^T$, with $\delta_e=B_a^{-1}\left(P_m+B_bP_L-B_cP_0\right)$ and $\omega_e=\1_{n_g}$. Otherwise, when DEOC is on, the equilibrium point is shifted to $x_{c}=[\delta_{c}~~\omega_{c}]^T$, with $\delta_{c}=\delta_e -B_a^{-1}B_c \Delta P$ and $\omega_{c}=\omega_e=\1_{n_g}$.

\subsubsection{Oscillation orbit} The system is initially in steady state at $x_e$. Now, assume the state variables drift away from $x_e$ due to a short-circuit. If $x_0=x(t_0)$ when the short circuit is cleared at time $t_0$, then the evolution of the state variables is given by $x(t)=M e^{\Lambda (t-t_0)}M^{-1} \left( x_0-x_e \right)+x_e, \forall~t \geq t_0$,
where $M=[q_1,q_2,...,q_i,...,q_{2 n_g}]$ (full rank), $q_i$ eigenvector related to $\lambda_i$, $\Lambda = M^{-1}A M=\diag \{\lambda_1,\lambda_2,...,\lambda_{2 n_g} \}$ (all distinct and different to zero). Considering $x(t)$ and its derivative $\dot{x}(t)=M\Lambda e^{\Lambda (t-t_0)}M^{-1} \left( x_0-x_e \right)$, the following hyperellipsoid in the plane $x$-$\dot{x}$ is obtained after some algebraic manipulations (black dashed line in Fig. \ref{fig:BangBangControl}):
{\small\begin{align}
(x-x_e)^T D (x-x_e)+&\dot{x}^T E \dot{x}=2(x_0-x_e)^T D (x_0-x_e)\label{eq:OscillationOrbit}
\end{align}}
where $D=\left( M^{-1} \right)^*M^{-1}$, $E=\left( M^{-1} \right)^*\left( \Lambda^{-1} \right)^*\Lambda^{-1}M^{-1}$, with $^*$ being the conjugate transpose. Note these matrices are real positive definite.
\subsubsection{DEOC activation} When the system moves along the oscillation orbit, DEOC is activated at time $t_{on}$. The equilibrium point is instantaneously shifted to $x_{c}$. The state variables are described by $x(t)=M e^{\Lambda (t-t_{st})}M^{-1} \left( x_{st}-x_{c} \right)+x_{c}$, $\forall~t \geq t_{on}$, with $x_{st}=x(t_{on})$. In the phase plane $x$-$\dot{x}$, this corresponds to (see red dot-dashed line in Fig. \ref{fig:BangBangControl}):
{\small\begin{align}
(x-x_{c})^T D (x-x_{c})+\dot{x}^T E \dot{x}=2(x_{st}-x_{c})^T D (x_{st}-x_{c})\label{eq:SwitchOnCondition}
\end{align}}
Note, having a different switching time $t_{on}$ would change $x_{st}$; as a result, this would change the amplitude of the hyperellipsoid centered at $x_c$---red dotted line in Fig.\ref{fig:BangBangControl}-(b).
\subsubsection{Switch-on time calculation}
To obtain an optimal $t_{on}$, the hyperellipsoid described by Eq. \eqref{eq:SwitchOnCondition} must contain the equilibrium point $x_e$. By setting $x=x_e$, then the unknown becomes the variable $x_{st}=x$. By ordering the resulting equation, and realizing that $\dot{x}=A(x_e-x_c)$ when the system reaches $x_e$, we obtain the following switching function:
\begin{figure}
\begin{center}
\includegraphics[width=\columnwidth]{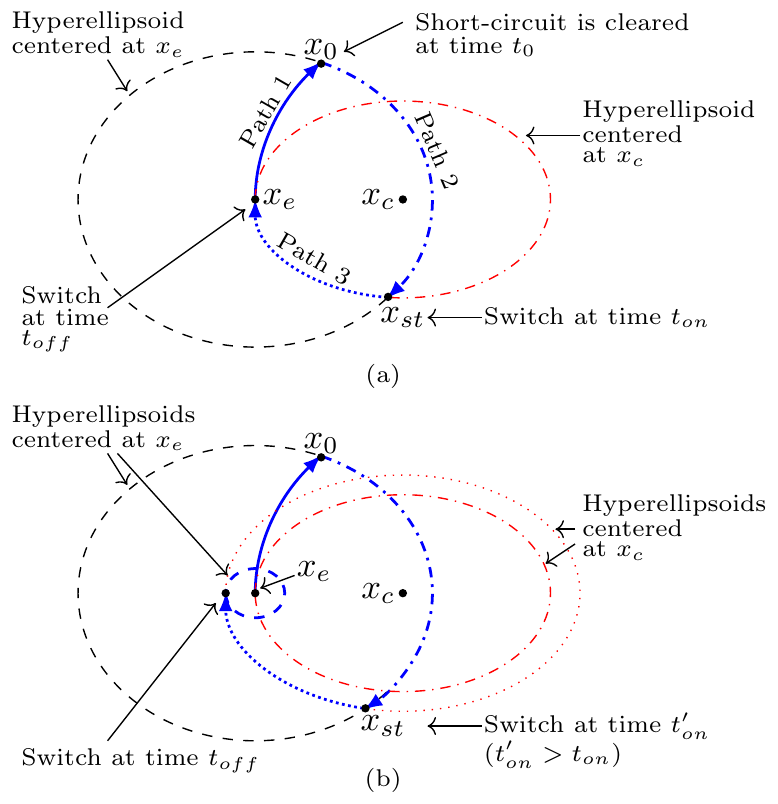}
\caption{Graphical description of the DEOC problem: (a) optimal trajectory, (b) sub-optimal trajectory}\label{fig:BangBangControl}
\end{center}
\end{figure}
\begin{align}
h(x)&=2(x_e-x_{c})^T D (x_e-x_{c})\hdots\nonumber\\
&-(x-x_{c})^T \left( D+A^T E A \right) (x-x_{c})\in \mathbb{R}\label{eq:OptimalSwitchingFunction}
\end{align}
Thus, while the system moves along the oscillation orbit (Path 2) with a vector $x$ ruled by Eq. \eqref{eq:OscillationOrbit}, DEOC must be turned on whenever the switching function vanishes, i.e., $h(x)=0$.
\subsubsection{Switch-off time calculation} In the ideal case, when DEOC is turned on at the exact optimal time $t_{on}$, DEOC must be switched off as soon as $x$ reaches $x_e$ (Path 3 in Fig. \ref{fig:BangBangControl}-a). In a more realistic case, however, it would be hard to switch DEOC on at the optimal time due to intrinsic delays and model errors, and the point $x_e$ may never be reached (Fig. \ref{fig:BangBangControl}-b). In this case, as the goal is to restrain oscillations after DEOC is turned off, the oscillation energy $E_k(t)=\omega_s (\omega-\1_{n_g})^TH(\omega-\1_{n_g})$  \cite{Silva2} can be used as a switching function to determine $t_{off}$. When $E_k(t)$ is reduced, the oscillation will be shrunk and $x$ will be confined to a closer orbit around $x_e$. Thus, when $x$ is orbiting around $x_c$ as DEOC is on, set $t_{off}=t$ if at a particular time $t$, $E_k(t)$ reaches a minimum.

\subsubsection{Applications}
\begin{figure}[t!]
\centering
\includegraphics[width=\columnwidth]{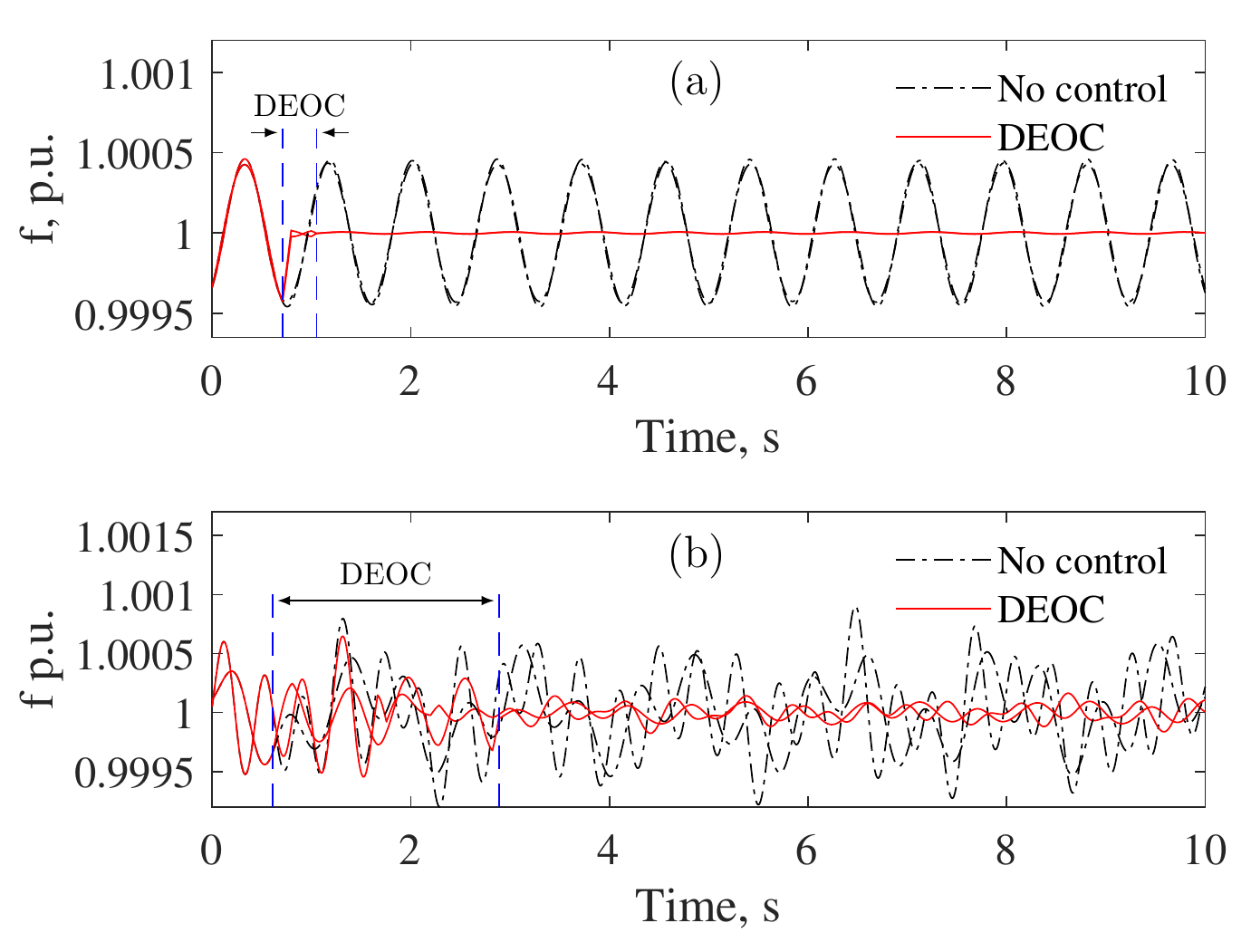}
\caption{DEOC application: (a) 9-bus test system, (b) 39-bus test system}
\label{fig:DEOC}
\end{figure}

This approach has been applied to a 9-bus and 39-bus test systems. Data has been obtained from reference \cite{Silva}. In both systems, SG1 has been defined as an infinite bus and a CC is considered in each non-generator bus. If the decision of how each CC would change its power output is taken in advance, then $\Delta P \in \mathbb{R}_{n_c}$ is defined, and the switching times $t_{on}$ and $t_{off}$ are determined as described above. If $\Delta P$ can be freely specified, a judicious definition of $\Delta P$ can restrain oscillations even further. In this letter, $\Delta P$ is defined such that the discrete actions cause the equilibrium point $x_c$ to lay only in the subspace related to the most significant excited oscillation modes. For the 9-bus test system, a short-circuit at bus 8 is considered. The first targeted eigenvalue is $\lambda_1 = j7.35$ rad/s and the power shift is defined as $\Delta P= -[16.8 \quad 32.4 \quad 26.6 \quad 62.1 \quad 54.9 \quad 44.7]^T$ MW. By applying the switching function $h(x)$ and by minimizing the oscillation energy afterward, the following switching times are obtained: $t_{on}=0.708$ s and $t_{off}=0.796$ s. For the second eigenvalue $\lambda_2 = j14.33$ rad/s, the following parameters have been determined: $\Delta P= -[18.4 \quad 35.6 \quad 29.1 \quad 68.1 \quad 60.1 \quad 49.0]^T$ MW, $t_{on}=1.000$ s and $t_{off}=1.051$ s. Overall, DEOC has been applied from times 0.708 s to 1.051 s, and the oscillations are reduced dramatically. The DEOC parameters have been determined for the 39-bus system in a similar fashion. Only 5 eigenvalues are targeted and DEOC is applied between times 0.607 s and 2.887 s. For legibility, the frequency of only a few SGs are shown in Fig. \ref{fig:DEOC}, with and without DEOC.

\subsubsection{Final remarks}
Discrete control can be applied to multi-modal systems for the first time, without the need for any model aggregation. Note that the goal is not to outperform traditional continuous control strategies,
but to enable a suitable control logic in EPT that recognizes their particular characteristics and limitations. Although the results validate the proposed scheme and show its effectiveness, further theoretical development is needed to: (a) determine optimal $\Delta P$ subject to CC restrictions, (b) use standard models for the grid and its components, e.g., SGs, PSS, communication network, that can better capture non-linearities and delays, (c) include stability analysis, and (d) update switching times based on these new models---calculated either through new switching functions or as a time correction based on the discrepancy of the idealized and more realistic representations.

\subsection{Discrete frequency excursion control (DFEC)}
Consider a SG and a synchronous motor connected through a loss-less short line. A CC is connected at the SG side. Both machines have a voltage controller, but only the SG has a governor (IEESGO model). The power setpoint of both machines is at 0.75 pu. The frequency evolution of the machines when the motor has an increase of 0.25 pu of its mechanical 
\begin{figure}
\centering
\includegraphics[width=\columnwidth]{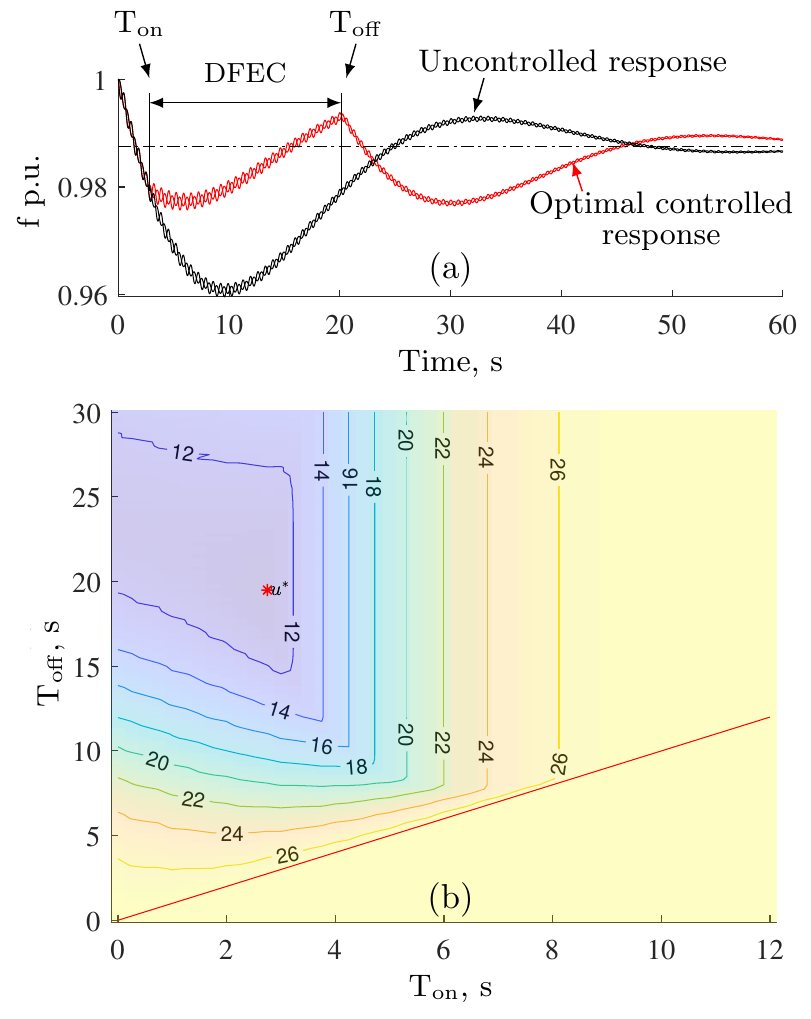}
\caption{DFEC after 0.25 pu mechanical power increase: (a) frequency, (b) $\Delta f_{\max}$ contour plot scaled by 1,000 when $\Delta P=0.1282$ pu }\label{fig:FreqExcursion}
\end{figure}
power is shown in Figure \ref{fig:FreqExcursion}-(a) (uncontrolled response, with a frequency nadir of 4\%). To investigate a discrete response, $\Delta P$, $T_{on}$ and $T_{off}$ are considered as optimization variables. Candidate solutions are evaluated using the cost function $\omega_{s.s.}-\min(\frac{\omega_1+\omega_2}{2})$ which measures the maximum excursion of the average frequency of both machines with respect to the steady-state value achieved by the system after the disturbance. An interior point optimization algorithm was employed to determine the optimal solution which is shown alongside the system response in Figure \ref{fig:FreqExcursion}-(a) (optimal controlled response, with a frequency nadir of about 2\%); the optimal variables are $\Delta P^{\ast}=0.1282$ pu, $T_{on}^{\ast}=2.77$ s and $T_{off}^{\ast}=22.25$ s. Additionally, a contour and level set plot shown in Figure \ref{fig:FreqExcursion}-(b), for the special case of $\Delta P=\Delta P^{\ast}$, was created by evaluating over the critical range of $T_{on}$ and $T_{off}$ values. These results shed light on the unique challenges faced in trying to design discrete control strategies for the frequency excursion problem. It demonstrates that for large power injections, the optimal control strategy is not to immediately offset the change in load with a corresponding change in the injected power, but rather to allow for the natural response of the speed governor to begin its response and inject power only during a critical window. An additional finding is that the optimal power injection window is not independent of the injected power magnitude, but rather, becomes more narrow as the injected power magnitude increases. A clear diminishing rate of return with respect to the maximum frequency excursion and length of the power injection window can also be seen. Intuitively, one can imagine that continuing to inject power well after the frequency excursion has reached its nadir does not continue to improve the maximum frequency excursion. Control strategies that continue injecting power unnecessarily in this manner are not only more inefficient in terms of power management, they are also less robust in that they waste reserves and will be less able to effectively mitigate a subsequent excursion event \cite{Briere}.

\bibliographystyle{IEEEtran}
\bibliography{References}

\end{document}